\documentclass[final]{svjour2}
\usepackage{graphicx}
\usepackage{wrapfig}
\usepackage{rotating}
\usepackage{amssymb}
\usepackage{bm}
\usepackage{mathptmx}
\usepackage[numbers]{natbib}
\makeatletter
\journalname{Journal of Low Temperature Physics}

\bibpunct{}{}{,}{s}{}{,}

\begin{document}

\newcommand{\hdblarrow}{H\makebox[0.9ex][l]{$\downdownarrows$}-}
\title{Two-superfluid Model of Two-component Bose-Einstein Condensates; First Sound and Second Sound}

\author{S. Ishino \and H. Takeuchi \and M. Tsubota }

\institute{Department of Physics, Osaka City University,Sumiyoshi-Ku, Osaka 558-8585, Japan\\
Tel.:+81-6-6605-2501\\ Fax:+81-6-6605-2522\\
\email{ishino@sci.osaka-cu.ac.jp}
}

\date{XX.06.2010}

\maketitle


\begin{abstract}
 Superfluid $^4$He at a finite temperature is described by the two-fluid model with the normal fluid component and the superfluid component.
 We formulate the two-fluid model for two-component BECs, namely two-superfluid model, starting from the coupled Gross-Pitaevskii equations.
 The two-superfluid model well corresponds to the two-fluid model in superfluid $^4$He.
 In a special condition, the two sound modes in the two-superfluid model behave like first and second sounds in the two-fluid model of superfluid $^4$He.
 
\end{abstract}

\section{Introduction}
Superfluid $^4$He has been thoroughly studied theoretically and experimentally in the field of low temperature physics since Kapitsa discovered superfluidity of $^4$He below the transition temperature\cite{Kapitsa}.
Tisza\cite{Tisza} and Landau\cite{Landau} succeeded in understanding the superfluidity of $^4$He with introducing the two-fluid model, which states that the system consists of normal fluid and superfluid being independent of each other and is described by
\begin{eqnarray}
\rho_{n} \Bigr(\frac{\partial{\bm v}_{n}}{\partial t}+({\bm v}_{n}\cdot\nabla){\bm v}_{n}\Bigl)&=&-\frac{\rho_{n}}{\rho}\nabla P-\rho_{n}\sigma\nabla T+\eta_{n}\nabla^2{\bm v}_{n},\\
 \rho_{s} \Bigr( \frac{\partial {\bm v}_{s}}{\partial t}+({\bm v}_{s}\cdot\nabla){\bm v}_{s}\Bigl)&=&-\frac{\rho_{s}}{\rho}\nabla P+\rho_{s}\sigma\nabla T.
 \end{eqnarray}
 Here $\rho _{n}$ and ${\bm v}_{n}$ are density and velocity of normal fluid, and $\rho _{s}$ and ${\bm v}_{s}$ are those of superfluid.
 $\sigma$ is entropy per unit mass of the normal fluid, $\rho =\rho _{n}+\rho _{s}$ is total density and $\eta _{n}$ is the coefficient of viscosity of the normal fluid.
The pressure gradient $\nabla P$ runs both components in the same direction and the thermal gradient $\nabla T$ does in the opposite direction.
Thermal counterflow driven by a thermal gradient is characteristic of superfluid $^4$He.
When the relative velocity between two components is large, they become dependent through the mutual friction ${\bm F}_{sn}$, which is added to Eqs. (1) and (2)\cite{Gorter}.
Other formulations for superfluid $^4$He are derived from the conservation law and equations of motion of mass density and entropy density.
The hydrodynamic equations\cite{Khalatnikov} are
\begin{eqnarray}
\frac{\partial\rho}{\partial t}&=&-\nabla\cdot{\bm j}, \\
\frac{\partial\sigma}{\partial t}&=&-\frac{\rho _{s}\sigma}{\rho}\nabla\cdot ({\bm v}_{n}-{\bm v}_{s}), \\
\frac{\partial{\bm j}}{\partial t}&=&-\nabla P, \\
\frac{\partial}{\partial t}({\bm v}_{n}-{\bm v}_{s})&=&-\frac{\rho\sigma}{\rho _{n}}\nabla T,
 \end{eqnarray}
 where ${\bm j}=\rho _{n}{\bm v}_{n}+\rho _{s}{\bm v}_{s}$.
 These equations yield the wave equation of mass density and entropy density, which leads to first sound and second sound.
 First sound is a mode of oscillation of total density and it exists generally in a usual fluid.
 While, second sound is a characteristic mode in superfluid $^4$He, in which entropy oscillates without oscillating total density, not existing in a usual fluid.

 An atomic BEC is one of the most important subjects in modern physics.
 Especially, two-component BECs are known to create various exotic structure of quantized vortices\cite{Kasamatsu05} and cause some characteristic hydrodynamic instability such as Kelvin-Helmholtz   instability\cite{Takeuchi10} and Rayleigh-Taylor instability\cite{Sasaki09}.
 In another paper, we investigate counterflow in two-component BECs which has many analogies with thermal counterflow in superfluid $^4$He.  
 For example, when the relative velocity exceeds a critical value, the counterflow becomes unstable and quantum turbulence appears like in thermal counterflow.
 In this work, we describe two-component BECs following the two-fluid model of superfluid $^4$He and obtain four equations similar to Eqs. (3)-(6).
 We derive two sound modes from these elementary equations.
 Two sound modes in two-component BECs were obtained by some other works\cite{Gladush}, but we have them correspond to first and second sounds.
 This is the main point of this work.
 Thus we can expect to improve interactive studies in superfluid $^4$He and two-component BECs with investigating common features between these.
\section{The two-fluid model in two-component BECs}

 We consider binary mixture of BECs described by the wave functions $\Psi _j=\sqrt{n_j}e^{i\phi_j}$ in the mean-field approximation at $T=0$ K, where the index $j$ refers to each component $j$ ($j=1,2$).
 The wave functions $\Psi _{j}$ are governed by the coupled Gross-Pitaevskii (GP) equations\cite{Pethick},
 \begin{eqnarray}
 i \hbar\frac{\partial}{\partial t}\Psi _{1}= - \frac{{\hbar}^{2}}{2m_{1}}\nabla^2\Psi _{1}+V({\bm r})\Psi _{1}+g_{11}|\Psi _{1}|^{2}\Psi _{1}+g_{12}|\Psi _{2}|^{2}\Psi _{1},\\
 i \hbar\frac{\partial}{\partial t}\Psi _{2}= - \frac{{\hbar}^{2}}{2m_{2}}\nabla^2\Psi _{2}+V({\bm r})\Psi _{2}+g_{22}|\Psi _{2}|^{2}\Psi _{2}+g_{12}|\Psi _{1}|^{2}\Psi _{2},
 \end{eqnarray}
 where $m_{j}$ is particle mass associated with the species, $g_{jj}$ is intracomponent interaction and $g_{12}$ is intercomponent  interaction.
We insert $\Psi _{j}$ into Eqs. (7) and (8) and obtain the hydrodynamic equations,
 \begin{eqnarray}
\frac{\partial\rho _{1}}{\partial t}&=&-\nabla\cdot (\rho _{1}{\bm v}_{1}), \\
\frac{\partial\rho _{2}}{\partial t}&=&-\nabla\cdot (\rho _{2}{\bm v}_{2}), \\
\rho _{1}\frac{\partial}{\partial t}{\bm v}_{1}&=&\rho _{1}\nabla\Bigl\{\frac{\hbar^2}{2m_{1}^2\sqrt{\rho _{1}}}\bigtriangleup\sqrt{\rho _{1}}-\frac{1}{2}v_{1}^2-\frac{1}{m_{1}^2}V\Bigr\}\nonumber\\
&\  &-\rho _{1}\nabla\Bigl(\frac{g_{11}}{m_{1}^2}\rho _{1}-\frac{g_{12}}{m_{1}m_{2}}\rho _{2}\Bigr), \\
\rho _{2}\frac{\partial}{\partial t}{\bm v}_{2}&=&\rho _{2}\nabla\Bigl\{\frac{\hbar^2}{2m_{2}^2\sqrt{\rho _{2}}}\bigtriangleup\sqrt{\rho _{2}}-\frac{1}{2}v_{2}^2-\frac{1}{m_{2}^2}V\Bigr\}\nonumber\\
&\  &-\rho _{2}\nabla\Bigl(\frac{g_{22}}{m_{2}^2}\rho _{2}-\frac{g_{12}}{m_{1}m_{2}}\rho _{1}\Bigr).
 \end{eqnarray}
 where $\rho _{j}=m_{j}n_{j}$ is mass density and ${\bm v}_{j}=\frac{\hbar}{m_{j}}\nabla\phi _{j}$ is superfluid velocity.
Equations (9) and (10) are equations of continuity for $\rho _{j}$ and Eqs. (11) and (12) are quasi-Euler equations for the superfluid velocity.

We will derive equations similar to Eqs. (1) and (2) to reveal correspondence between superfluid $^4$He and two-component BECs.
Here we consider a uniform system and apply the long-wavelength approximation, so the potential term and the quantum pressure term in Eqs. (11) and (12) are neglected.
The pressure of the whole system is
 \begin{eqnarray}
 {\tilde P}=\frac{g_{11}\rho _{1}^2}{2m_{1}^2}+\frac{g_{22}\rho _{2}^2}{2m_{2}^2}+\frac{g_{12}\rho_{1}\rho_{2}}{m_{1}m_{2}},\nonumber
 \end{eqnarray}
 since $P=-\partial E/\partial V$.
 Then Eqs. (11) and (12) turn into
  \begin{eqnarray}
 \rho _{1} \Bigr( \frac{\partial {\bm v}_{1}}{\partial t}+({\bm v}_{1}\cdot\nabla){\bm v}_{1}\Bigl)=-\frac{1}{2}\nabla {\tilde P}-\frac{1}{2} {\tilde \nabla}{\tilde T},\\
 \rho _{2} \Bigr( \frac{\partial {\bm v}_{2}}{\partial t}+({\bm v}_{2}\cdot\nabla){\bm v}_{2}\Bigl)=-\frac{1}{2}\nabla {\tilde P}+\frac{1}{2} {\tilde \nabla}{\tilde T},
 \end{eqnarray}
 with
  \begin{eqnarray}
 {\tilde \nabla}{\tilde T}=\frac{g_{11}}{2m_{1}^2}\nabla\rho _{1}^2-\frac{g_{22}}{2m_{2}^2}\nabla\rho _{2}^2+\frac{g_{12}}{m_{1}m_{2}}(\rho _{1}\nabla\rho _{2}-\rho _{2}\nabla\rho _{1}).\nonumber
 \end{eqnarray}
 It is impossible to describe the right hand side by gradient of some scalar potential because $\rho _{1}$ and $\rho _{2}$ are spatially dependent, but we represent it by ${\tilde \nabla}{\tilde T}$ in order to emphasize the correspondence to $\nabla T$ in Eqs. (1) and (2).
 From Eqs. (13) and (14), we can find that two-component BECs are driven by two terms.
The pressure gradient $\nabla{\tilde P}$ runs both components in the same direction, while ${\tilde \nabla}{\tilde T}$ runs them oppositely.
This nature is just the same as one of the two-fluid model in superfluid $^4$He.

\section{First sound and second sound in two-component BECs}
In this section, we will derive two sound modes from the four elementary equations in two-component BECs and let them correspond to first and second sounds.
Here we assume that superfluid velocities ${\bm v}_{j}$ are small and the non-linear terms are neglected.
By making Eq. (9) $\pm$ Eq. (10) and Eq. (11) $\pm$ Eq. (12) we obtain 
 \begin{eqnarray}
\frac{\partial}{\partial t}\rho _{+}&=&-\nabla\cdot{\bm j}_{+}, \\
\frac{\partial}{\partial t}\rho _{-}&=&-\nabla\cdot{\bm j}_{-}, \\
\frac{\partial}{\partial t}{\bm j}_{+}&=&-\nabla\Bigl\{\frac{g_{11}}{8m_{1}^2}(\rho _{+}+\rho _{-})^2+\frac{g_{22}}{8m_{2}^2}(\rho _{+}-\rho _{-})^2+\frac{g_{12}}{4m_{1}m_{2}}(\rho_{+}^2-\rho_{-}^2)\Bigr\}, \\
\frac{\partial}{\partial t}{\bm j}_{-}&=&-\nabla\Bigl\{\frac{g_{11}}{8m_{1}^2}(\rho _{+}+\rho _{-})^2-\frac{g_{22}}{8m_{2}^2}(\rho _{+}-\rho _{-})^2\Bigr\}\nonumber\\
                                                               &\  & +\frac{g_{12}}{4m_{1}m_{2}}\Bigl\{(\rho _{+}+\rho _{-})\nabla(\rho _{+}-\rho _{-})-(\rho _{+}-\rho _{-})\nabla(\rho _{+}+\rho _{-})\Bigr\},
 \end{eqnarray}
 where $\rho _{\pm}\equiv\rho _{1}\pm\rho _{2}$ and ${\bm j}_{\pm}\equiv\rho _{1}{\bm v}_{1}\pm\rho _{2}{\bm v}_{2}$.
Because first sound means oscillation of $\rho$ with two components in phase and second sound means oscillation of $\sigma$ with them out of phase, we expect that oscillations of $\rho _{+}$ and $\rho _{-}$ correspond respectively to first and second sounds.
 Now the right hand sides of Eqs. (17) and (18) should be $\nabla{\tilde P}$ and ${\tilde \nabla}{\tilde T}$ respectively.
We can write $\nabla {\tilde P}$ and ${\tilde \nabla}{\tilde T}$ as functional of $\rho _{+}$ and $\rho _{-}$ by
\begin{eqnarray}
 \nabla {\tilde P}&=&{\rm A}\nabla \rho _{+}+{\rm B}\nabla \rho _{-},\\
 {\tilde \nabla}{\tilde T}&=&{\rm C}\nabla \rho _{+}+{\rm D}\nabla \rho _{-},
\end{eqnarray}
where
\begin{eqnarray}
  {\rm A}&=&\frac{g_{11}}{4m_{1}^2}(\rho _{+}+\rho _{-})+\frac{g_{22}}{4m_{2}^2}(\rho_{+}-\rho_{-})+\frac{g_{12}}{2m_{1}m_{2}}\rho_{+},\nonumber\\
  {\rm B}&=&\frac{g_{11}}{4m_{1}^2}(\rho _{+}+\rho _{-})-\frac{g_{22}}{4m_{2}^2}(\rho_{+}-\rho_{-})-\frac{g_{12}}{2m_{1}m_{2}}\rho_{-},\nonumber\\
  {\rm C}&=&\frac{g_{11}}{4m_{1}^2}(\rho _{+}+\rho _{-})-\frac{g_{22}}{4m_{2}^2}(\rho_{+}-\rho_{-})+\frac{g_{12}}{2m_{1}m_{2}}\rho_{-},\nonumber\\
  {\rm D}&=&\frac{g_{11}}{4m_{1}^2}(\rho _{+}+\rho _{-})+\frac{g_{22}}{4m_{2}^2}(\rho_{+}-\rho_{-})-\frac{g_{12}}{2m_{1}m_{2}}\rho_{+}.\nonumber
\end{eqnarray}
The wave equations derived from Eq.(15)-(18) are reduced to
\begin{eqnarray}
\frac{\partial^2 }{\partial t^2}\rho_{+}={\rm A}\nabla^2\rho_{+}+{\rm B}\nabla^2\rho_{-},\\
\frac{\partial^2 }{\partial t^2}\rho_{-}={\rm C}\nabla^2\rho_{+}+{\rm D}\nabla^2\rho_{-}.
\end{eqnarray}
Considering the plane waves that $\rho _{+}$ and $\rho _{-}$ oscillate around the equilibrium values $\rho _{+}^0$ and $\rho _{-}^0$ with the frequency $\omega$ and the wave number ${\bm k}$ like
\begin{eqnarray}
\rho _{+}=\rho _{+}^0+\delta\rho _{+}{\rm exp}[i({\bm k}\cdot{\bm r}-\omega t)],\nonumber\\
\rho _{-}=\rho _{-}^0+\delta\rho _{-}{\rm exp}[i({\bm k}\cdot{\bm r}-\omega t)],\nonumber
\end{eqnarray}
the sound velocities are
 \begin{eqnarray}
 c^2&=&\frac{g_{11}}{4m_{1}^2}(\rho _{+}^0+\rho _{-}^0)+\frac{g_{22}}{4m_{2}^2}(\rho _{+}^0-\rho _{-}^0)\nonumber\\
 &\  &\  \  \pm\sqrt{\Bigl\{\frac{g_{11}}{4m_{1}^2}(\rho _{+}^0+\rho _{-}^0)-\frac{g_{22}}{4m_{2}^2}(\rho _{+}^0-\rho _{-}^0)\Bigl\}^2+\frac{g_{12}^2}{4m_{1}^2m_{2}^2}(\rho _{+}^{0^{2}}-\rho _{-}^{0^{2}})},
\end{eqnarray}
where $c\equiv\omega/|{\bm k}|$.
These are obtained from the dispersion relation of the  Bogoliubov excitations\cite{Gladush, Pethick} in two-component BECs in the limit of long wavelength.

In superfluid $^4$He first and second sounds are modes that $\rho$ and $\sigma$ independently oscillate. 
However, Eq. (23) does not necessarily describe first and second sounds because two modes are mixed.
We can find that $\rho _{+}$ and $\rho _{-}$ oscillate independently when ${\rm B}$ and ${\rm C}$ vanish in Eqs. (21) and (22).
This conditions are reduced to
\begin{eqnarray}
\rho _{1}^0&=&\rho _{2}^0,\nonumber\\
\frac{g_{11}}{m_{1}^2}&=&\frac{g_{22}}{m_{2}^2}.\nonumber
\end{eqnarray}
Then sound velocities of the two modes are
 \begin{eqnarray}
 c_{\pm}^2&=&s^2\pm\frac{g_{12}\rho^0}{m_{1}m_{2}},
\end{eqnarray}
where $s=\sqrt{g_{11}\rho _{1}^0/m_{1}^2}=\sqrt{g_{22}\rho _{2}^0/m_{2}^2}$ and $\rho^0=\rho _{1}^0=\rho _{2}^0$.
The mode of $c_{+}^2$ is oscillation of $\rho _{+}$, first sound, and the mode of $c_{-}^2$ is oscillation of $\rho _{-}$, second sound.
First sound velocity increases with $g_{12}$ and second sound velocity decreases with $g_{12}$ (Fig.1).
When $|g_{12}|>g\equiv\sqrt{g_{11}g_{22}}$, $c_{+}$ or $c_{-}$ becomes imaginary so that the dynamical instability leads to the collapse or the phase separation in the two-component BECs\cite{Pethick}.
\begin{figure}
\centering
 \includegraphics[width=6cm]{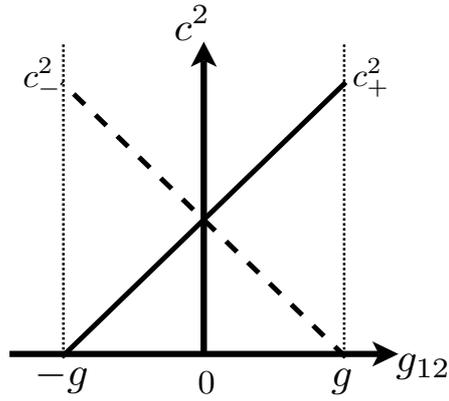}
 \caption{
 Velocity of first and second sounds as a function of $g_{12}$.
 The solid and dashed line refers to first sound $c_{+}^2$ and second sound $c_{-}^2$ respectively.
 }
\end{figure}%

\section{Summary}
We formulated the two-fluid model for two-component BECs, starting from the coupled GP equations.
This model well corresponds to the two-fluid model in superfluid $^4$He expect for the mutual friction term.
We obtained the condition that two sound modes are independent of each other like first and second sounds in superfluid $^4$He.
Second sound has an important role to investigate quantum turbulence in superfluid $^4$He.
We are interested in how second sound interacts with a vortex in two-component BECs.
In the future, we should investigate "mutual friction" induced by the interaction between vortices and the Bogoliubov excitations in two-component BECs.
The details will be reported soon elsewhere.

\begin{acknowledgements}
 M.T. acknowledges the support of a Grant-in-Aid for Scientific Research from JSPS (Grant No. 21340104).
\end{acknowledgements}

\end{document}